\documentclass[11pt,draft,aps,prb,showpacs]{revtex4}

\usepackage{epsf}
\usepackage{dcolumn}

\begin{document}

\def\mc  {\multicolumn}
\def\tn  {\tablenote}

\title {On the observed irregular melting temperatures of
         free sodium clusters}

\author{S. Chacko and D. G. Kanhere}
\affiliation{
Centre for Modeling and Simulation, and Department of Physics,
University of Pune, Ganeshkhind, Pune 411 007, India.}
\author{S. A. Blundell}
\affiliation{
D\'epartement de Recherche Fondamentale sur la
Mati\`ere Condens\'ee, CEA--Grenoble/DSM \\
17 rue des Martyrs, F--38054 Grenoble Cedex 9, France.}

\begin{abstract}
Density--functional simulations have been performed on Na$_{55}$,
Na$_{92}$ and Na$_{142}$ clusters in order to understand the
experimentally observed melting properties 
[M. Schmidt \textit{et al.}, Nature~(London) \textbf{393}, 238~(1998)].
The calculated melting temperatures are in excellent agreement with the
experimental ones. The calculations reveal a rather subtle interplay
between geometric and electronic shell effects, and bring out the fact
that the quantum mechanical description of the metallic bonding is crucial
for understanding quantitatively the variation in melting temperatures
observed experimentally.
\end{abstract}


\maketitle

\section{Introduction}

There is currently great interest in the statistical mechanics of
finite--sized versions of systems that display phase transitions in
the infinite--size limit.  Recently, atomic clusters have provided an
example of a finite system whose caloric curve may be measured,
yielding evidence for a ``melting'' transition over a broad range of
temperatures.\cite{haberland0,haberlandnegCV,haberlandCR,jarroldGa}
The clusters involved in these experiments were both small~(containing
from a few atoms to several hundreds of atoms) and free~(not supported
on a surface).  Now, according to old thermodynamic
models,\cite{pawlow,buffat} a small particle should melt at a lower
temperature than the bulk because of the effect of the surface, with
the reduction in melting temperature being roughly proportional to
$1/R$, where $R$ is the radius of the particle.  This effect has been
verified quantitatively for particles of mesoscopic size~(upwards of
several thousand atoms) supported on a surface.\cite{buffat,boiko}
More recently, however, in a series of experiments on free Na clusters
in the size range 55 to 350, Haberland and
co--workers\cite{haberland0,haberlandnegCV,haberlandCR} found that the
simple $1/R$ scaling is lost; they did observe a substantial
lowering~(by about 30\%) of the melting temperature compared to bulk
sodium, but accompanied by rather large size--dependent fluctuations
in the melting temperatures.  In spite of considerable theoretical
effort,\cite{manninen99,manninen04,cs,bb01,aguado,aguado-nalarge} the
precise form of the observed fluctuations remains unexplained to date.

As is well known, metallic clusters~(such as sodium) possess the same
first few ``magic'' sizes as atomic nuclei, $N=2$, 8, 20, \ldots,\cite
{knight84} corresponding to particularly stable systems in which the
delocalized valence electrons form closed fermionic shells. Unlike
nuclei, however, metallic clusters also contain positively charged
ions, which are much heavier than electrons and may be treated
classically to a good approximation, leading to the possibility that
geometric packing effects may enter into competition with electronic
shell effects in determining certain properties.  The precise pattern
of melting points observed experimentally\cite{haberland0,haberlandCR}
shows maxima at sizes that correspond neither to magic numbers of
valence electrons nor to complete Mackay icosahedra of ions, thereby
suggesting a subtle interplay between geometric and quantum electronic
effects.\cite{haberland0,haberlandCR,manninen99,manninen04}

Reliable simulations to determine the melting properties of clusters
are made difficult by a combination of two crucial requirements: the
need to compute the ionic potential--energy surface accurately, and
the need for high statistics to converge the features in the caloric
curve.  Good statistics have been obtained using a variety of
\emph{parametrized} interatomic potentials,\cite{manninen04,cs} as
well as with a treatment of the valence electrons in the extended
Thomas--Fermi~(ETF)\cite{bb01} and related density--based~(DB)
approximations.\cite{aguado,abhijat01} But these attempts have failed
to reproduce crucial features of the experimental results, such as the
precise pattern of observed melting temperatures.  An interesting
observation is that in all earlier simulations the melting temperature
of Na$_{55}$ has been considerably underestimated~(by about 100~K
lower than the experimental value).

Clearly, what is required is a more realistic treatment of interacting
electrons, in particular, one that incorporates correctly electronic
shell effects, which are explicitly excluded from the above--mentioned
work using parametrized, ETF, or DB potentials.  We have recently
demonstrated the power of \emph{ab initio} density--functional
theory~(DFT), in the Kohn--Sham (KS) formulation, for simulating the
melting properties of small Sn\cite{snmelting} and Ga\cite {gamelting}
clusters in the size range 10$-$20.  Unfortunately, the computational
expense of the KS approach, combined with the relatively large sizes
of the Na data~($N=$ 55--350), have so far made it difficult to
perform KS simulations with high statistics in the size range of the
experiment.  Recently, a KS simulation of
Na$_{55}{}^{+}$,\cite{manninen99,manninen04} performed with limited
statistics, shows encouraging results, with an estimated melting
temperature $T_m$ between 300 and 350~K [$T_m(\mathrm{expt})\approx
290$~K for Na$_{55}{}^{+}$].

In this paper, we report the first \emph{ab initio} KS thermodynamic
calculations up to size $N=$ 142, within the size range of the
experiment.  We present simulations for Na$_{N}$ with $N=$ 55, 92, and
142, each with total sampling times of the order of 2--3~ns.  For each
size there is a pronounced peak in specific--heat curve, whose
position agrees with the experimental finding to better than 8\%.  
This error is also the approximate level of statistical error for the
simulation times used, suggesting that a KS approach is capable of
quantitative predictions of melting temperatures at this level of
accuracy or better.  Analysis of the structural and dynamical
information in these simulations sheds new light on the respective
roles of geometric and electronic effects in this intriguing problem.

\section{Computational Details}

\begin{figure}
{\centerline {\epsfxsize=1.0\textwidth  \epsffile{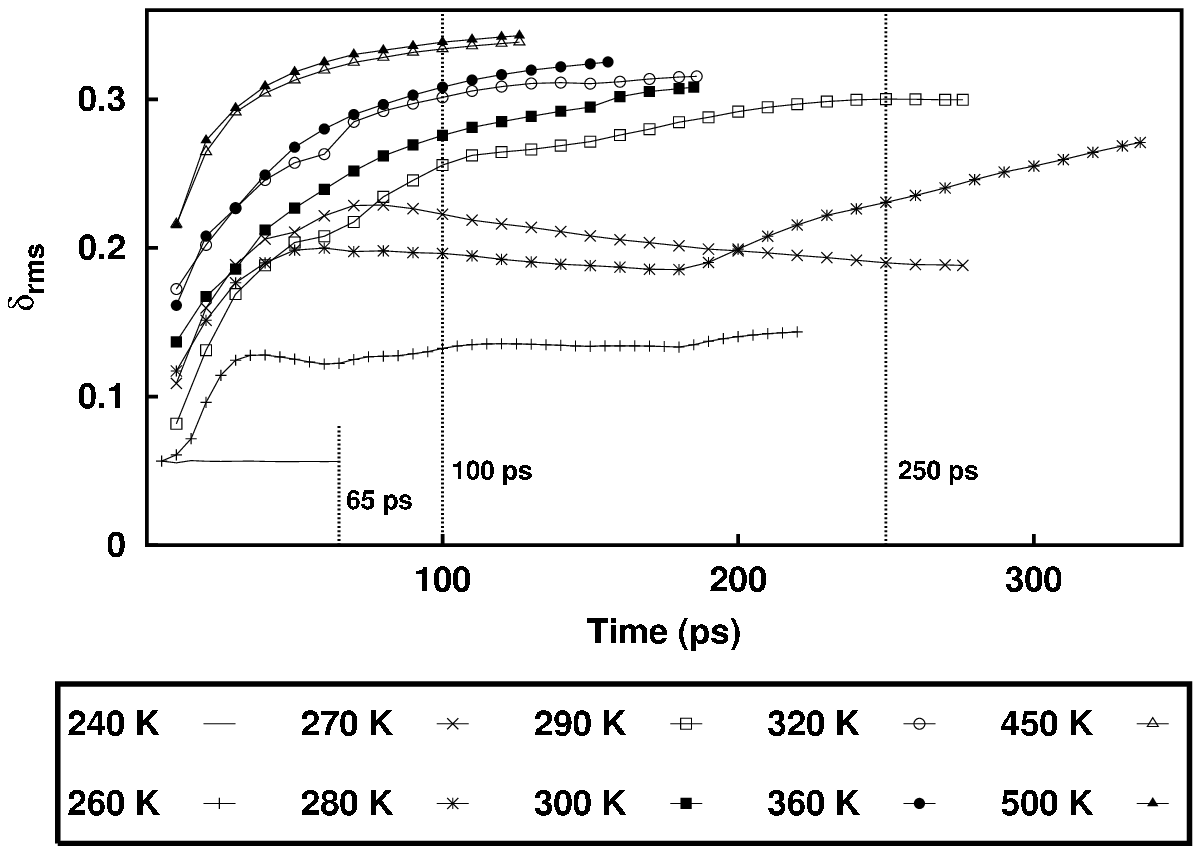}}}
{\caption{
\label{fig.na55.delta_conv}
The root--mean--square bond--length fluctuations $\delta_{\rm rms}$
for Na$_{55}$, as a function of the total time used to compute the
average value.
%
%
}}
\end{figure}

Our simulations have been carried out using isokinetic
Born--Oppenheimer molecular--dynamics with ultrasoft pseudopotentials
in the local density approximation, as implemented in the
\textsc{vasp} package.\cite{vasp} We have carried out calculations for
up to 17 temperatures for Na$_{55}$ and 14 temperatures for Na$_{92}$
and Na$_{142}$ in the temperature range 100 to 500~K. The simulation
times are 90~ps for temperatures below
$T_m$, and between 150--350~ps for temperatures near the melting
transition or higher. Such high statistics have to be performed in
order to achieve good convergence of thermodynamic indicators such as
the root-mean-square bond-length fluctuation
$\delta_{\rm rms}$.\cite{deltarms} In Fig.~\ref{fig.na55.delta_conv},
we show $\delta_{\rm rms}$ for Na$_{55}$ as a function of the total
time used to compute the time averages.  We have discarded at least
the first 25~ps at each temperature to allow for thermalization. This
plot shows that, for lower temperatures (below the transition
temperature, that is, $T\le240$~K) simulation times of the order of
60--70~ps are sufficient for convergence. (For a discussion of melting
temperatures $T_m$, see Sec.~\ref{results-discussion}.)  For the higher
temperatures (300~K and above), at least 100--300~ps are required to
achieve proper convergence.  However, near the melting transition,
namely, around 270 and 280~K, $\delta_{\rm rms}$ is rather poorly
converged even up to about 300~ps, indicating the need for higher
statistics at these temperatures.  Further, with increase in the
cluster size, these averaging times may increase, making the \emph{ab
initio} simulations of larger clusters very difficult.

An energy cutoff of 3.6~Ry is used
for the plane--wave expansion of the wavefunction, with a convergence
in the total energy of the order of 0.0001--0.0005~eV.
Increasing the energy cutoff to, say, about 22~Ry causes a change of
only 0.31\% and 0.28\%, respectively, in the binding energy and
bond--length of the Na$_{2}$ dimer. We have also checked these results
against those obtained by the projected augmented wave~(PAW)
method,\cite{paw} taking only the $1s^2$ electrons in the core and
with an energy cutoff of about 51~Ry. The Na$_{2}$ binding energy and
bond--length according to the present DFT method differ by less than
3\% from those given by the PAW method (the PAW method yielding
smaller energies and bond--lengths). The DFT method employed for our
thermodynamic simulations can therefore be considered to be of
comparable accuracy to the (nearly) all--electron method.  The lengths
of the simulation cells for Na$_{55}$, Na$_{92}$, and Na$_{142}$ are
24, 28, and 31~\AA, respectively.

\begin{figure}
{\centerline {\epsfxsize=1.0\textwidth  \epsffile{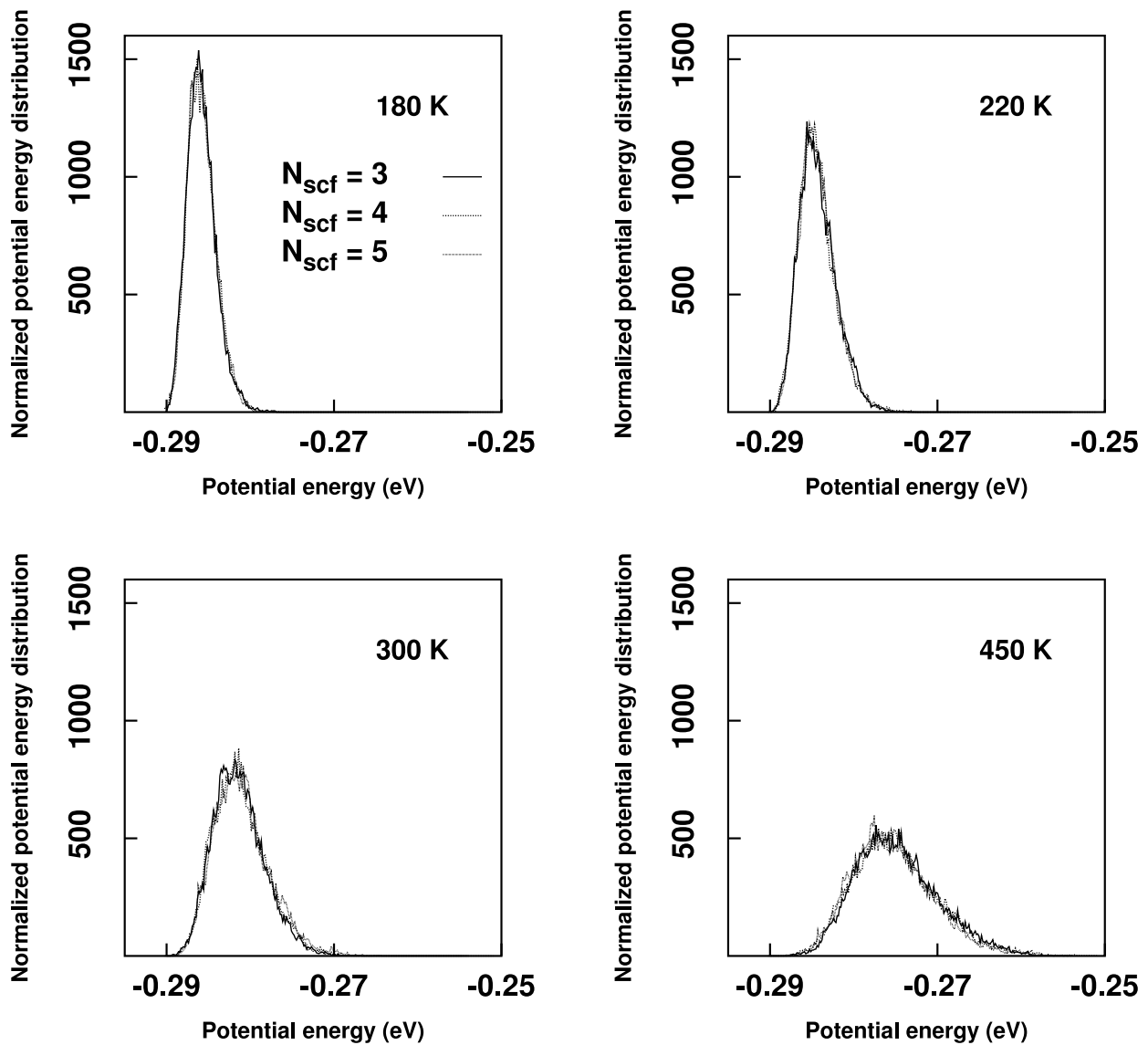}}}
{\caption{
\label{fig.na8.hist}
Comparison of the potential--energy distributions of the Na$_8$
cluster at 180, 220, 300, and 450~K, computed with minimum $\rm N_{\rm
scf}=$ 3, 4, 5, iterations\cite{vasp-wfn-extrapol} used for obtaining
the self--consistent field solution of the Kohn--Sham equations.
}}
\end{figure}

We have also performed certain tests on the Na$_8$ cluster to check
the convergence of the total energy and forces at high temperatures.
In Fig.~\ref{fig.na8.hist}, we plot the histograms of
potential--energy for the Na$_8$ cluster at 180, 220, 300, and 450~K
with a minimum of 3, 4, and 5 iterations\cite{vasp-wfn-extrapol} for
obtaining the self--consistent field~(SCF) solution of the KS
equations. Clearly, all the plots overlap, indicating that about 4--6
SCF iterations are sufficient to obtain a good convergence of the
potential--energy. The force convergence test has been performed on
the Na$_8$ cluster at 400~K. The forces are converged up to about
0.0001~eV/{\AA} in just 4 SCF iterations. Hence, in all our
calculations, we have used a minimum of 4, and in some cases up to 5
or 6 iterations for self--consistency.

The resulting trajectory data have been used to compute the ionic
specific--heat, via a multihistogram fit, as well as other
thermodynamic indicators.  Details can be found in
Ref.~\onlinecite{abhijat01}.

\section{Results and Discussion}
\label{results-discussion}

\begin{figure}
{\centerline {\epsfxsize=0.9\textwidth  \epsffile{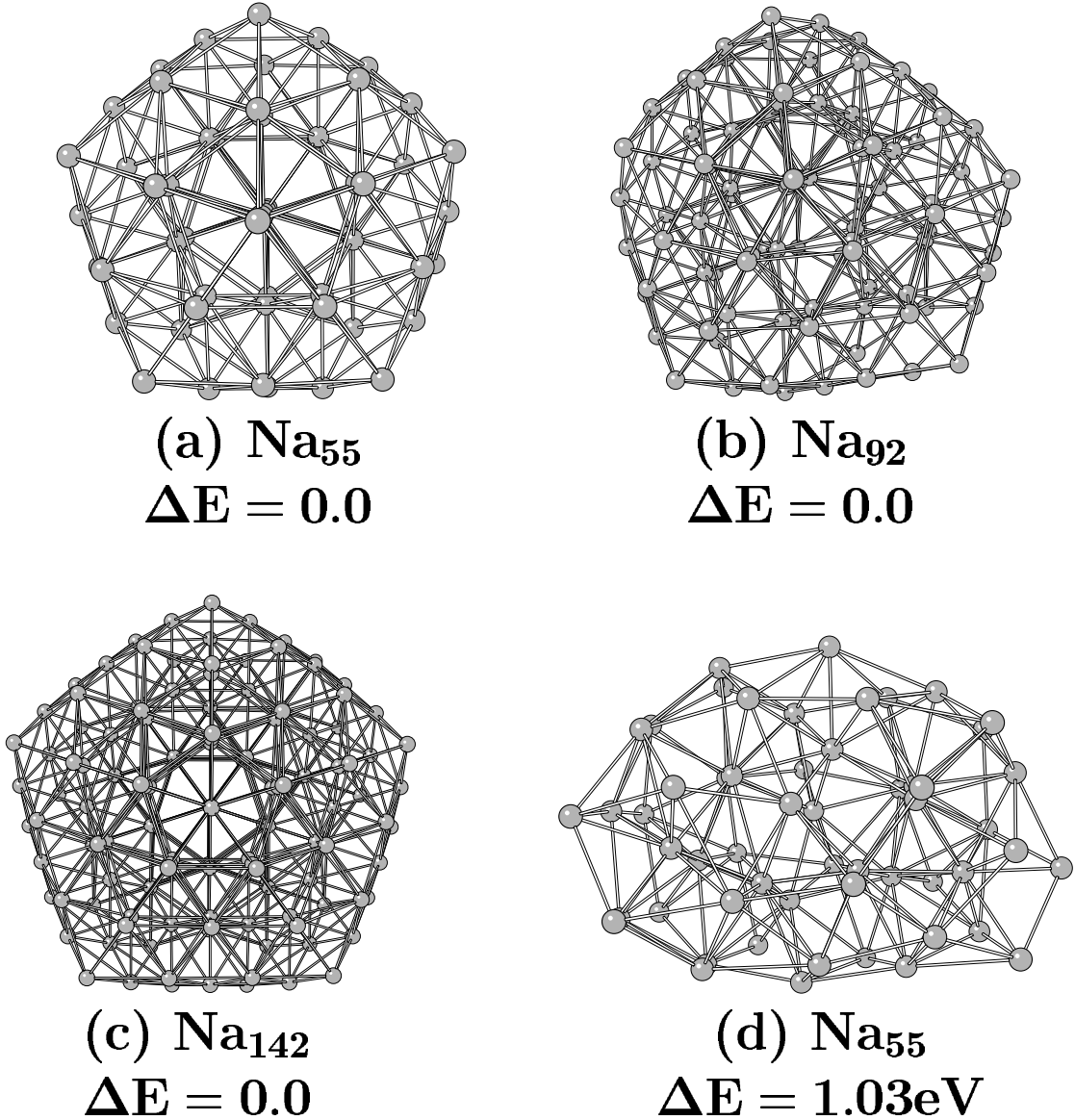}}}
\caption{
\label{fig.struc}
    Ground state geometries of (a)~Na$_{55}$, (b)~Na$_{92}$,
    and (c)~Na$_{142}$. Structure (d)~is a representative deformed
    excited--state structure of Na$_{55}$.
}
\end{figure}

We have carried out a substantial search for the lowest--energy ionic
structures.  A basin--hopping algorithm\cite{bhop} with around
10$^{5}$ basin hops was employed to generate several hundred
structures using the second--moment approximation~(SMA) parametrized
interatomic potential of Li \textit{et al.}\cite{smali}.  About 25 of
the lowest of these structures were then used as input for relaxation
within the KS approach.  In addition, several quenches were carried
out for structures selected from a long high--temperature \emph{ab
initio} MD run, typically taken at a temperature well above the
melting point~(say, for $T>400$~K). In this way, we have obtained
about 20--25 distinct geometries for each size.  Most of these
structures should be close to the true ground--state, although even
this method may not yield the correct absolute ground--state
structure. The resulting lowest--energy structures found for the three
sizes are shown in Figs.~\ref{fig.struc}(a)--\ref{fig.struc}(c).  The
lowest--energy geometry of Na$_{55}$ is found to be a two--shell
Mackay icosahedron (with a very slight distortion from perfect
regularity), in conformity with previous theory\cite{manninen99,brack}
and experimental evidence.\cite{icoexpt} Na$_{142}$ is also found to
be icosahedral, with 5 atoms missing from the outermost shell.  (A
three--shell Mackay icosahedron has 147 atoms.) As to Na$_{92}$, the
structure consists of surface growth over a slightly distorted
icosahedral Na$_{55}$ core.  The additional 37 surface atoms are
accommodated so as to maintain a roughly spherical shape.  Note that
for Na$_{92}$, the SMA basin--hopping process has an unclear
convergence with respect to the exact configuration of these surface
atoms, and so even for the SMA potential we are unlikely to have found
the true global ground--state structure.  However, since our
lowest--energy SMA structure turns out to be quite spherical, and
since Na$_{92}$ has a closed--shell configuration of valence
electrons, we believe that the overall shape of our lowest--energy KS
structure is likely to be correct.

\begin{figure}
{\centerline {\epsfxsize=0.9\textwidth  \epsffile{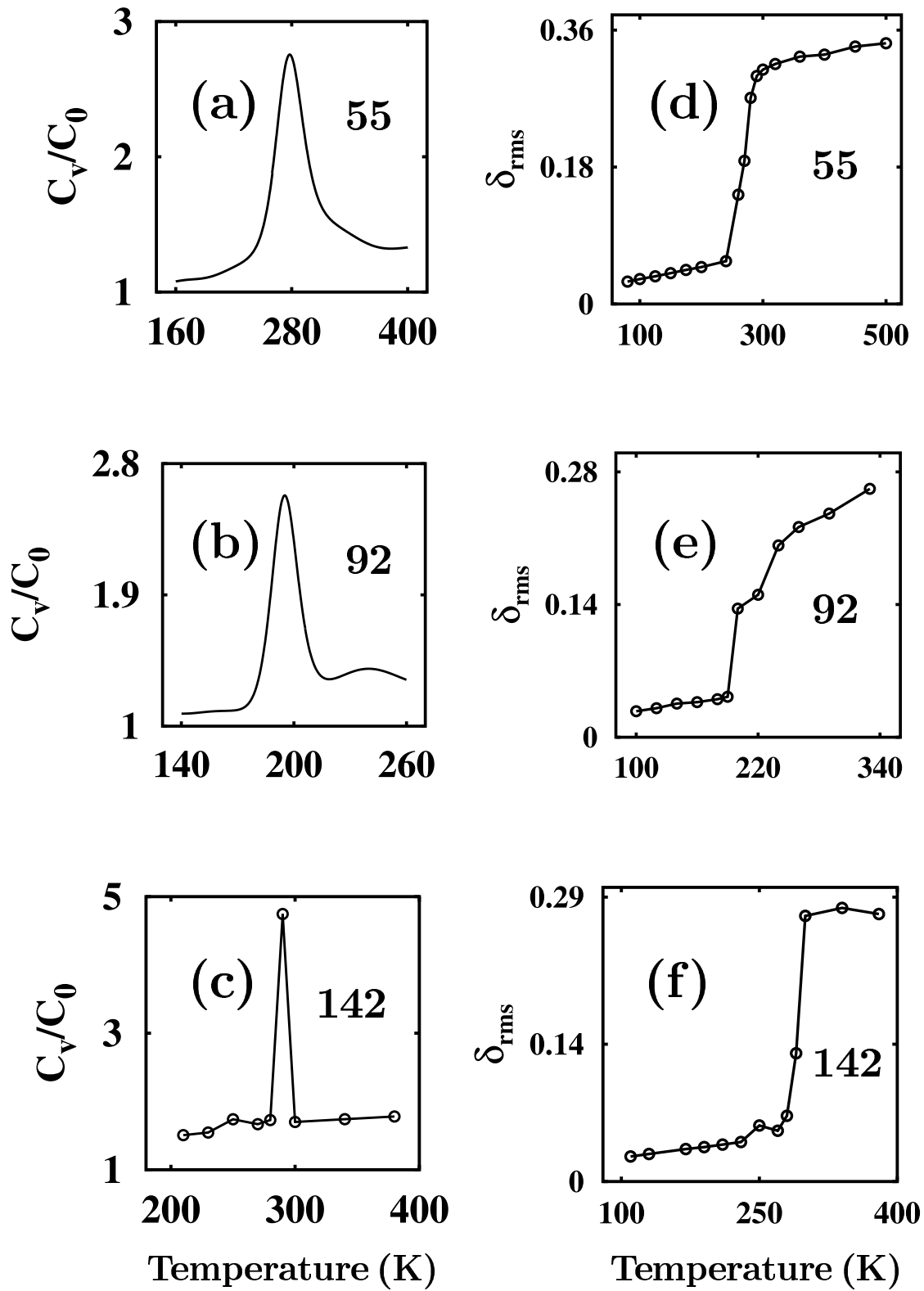}}}
{\caption{
\label{fig.spheat_delta}
     Left panel:
     Normalized canonical specific heat for
     (a)~Na$_{55}$, (b)~Na$_{92}$, and (c)~Na$_{142}$.
     $C_0=(3N-9/2)k_B$ is the zero--temperature classical
     limit of the rotational plus vibrational canonical
     specific heat.
     Right panel:
     Root--mean--square bond--length fluctuation
     $\delta_{\rm rms}$\protect\cite{deltarms} for
     (d)~Na$_{55}$, (e)~Na$_{92}$, and (f)~Na$_{142}$.
} }
\end{figure}

Next we examine the specific--heat for the three clusters
investigated, which are plotted in
Figs.~\ref{fig.spheat_delta}(a)--\ref{fig.spheat_delta}(c).  These
curves feature a single dominant peak with a width varying from about
40~K for Na$_{55}$ to about 20~K for Na$_{142}$, in general agreement
with the experimental observation.  A detailed investigation of the
ionic dynamics shows this peak to correspond to a ``melting--like''
transition, from a low--temperature solidlike state, in which ions
vibrate about fixed points~(and the cluster as a whole may rotate), to
a higher--temperature liquidlike state with a diffusive ionic
dynamics.  As one illustration of this, we show in
Figs.~\ref{fig.spheat_delta}(d)--\ref{fig.spheat_delta}(f) the
root--mean--square bond--length fluctuation
$\delta_{\mathrm{rms}}$.\cite{deltarms} For each size there is a sharp
step in $\delta_{\mathrm{rms}}$ that correlates closely with the peak
in the specific--heat.  The ``melting'' phenomenology found is
qualitatively similar to that found in earlier studies with simpler
potentials.\cite{cs,bb01,aguado-nalarge}

\begin{table}
\caption
{
    \label{table.data}
    Melting temperatures, latent heat, and $\delta E$~(see text) for
    Na$_{N}$ given by the present Kohn--Sham~(KS) approach, the
    second--moment approximation~(SMA) potential, and a
    density--based~(DB) approach.  The statistical error in the KS
    melting 
    temperatures is about 8\%.
}
\begin{ruledtabular}
\begin{tabular}{cccccc}
     &  \mc{5}{c}{Melting Temperature~(K)} \\
$N$  &  KS  &  Expt.\footnotemark[1]  &  SMA\footnotemark[2]  &  
SMA\footnotemark[3]  &  DB\footnotemark[4]  \\
\hline
55   & 280   &  290  & 175   & 162   & 190  \\
92   & 195   &  210  & (170)\footnotemark[5] & (133)\footnotemark[5] & 240  \\
142  & 290   &  270  & 240   &  186  & 270  \\
\hline
     &  \mc{4}{c}{Latent Heat (meV/atom)}  & $\delta E$ (meV/atom)\\
$N$  &  \mbox{KS}  &  \mbox{Expt.}\footnotemark[1]  & 
\mbox{SMA}\footnotemark[2]  &  \mbox{SMA}\footnotemark[3] & \mbox{KS}\\
\hline
55   &  13.8 & 13.0 & 20.1 & 8.3  &  53.2 \\
92   &  6.4 & 6.0 & (16.9)\footnotemark[5] & (4.2)\footnotemark[5] & 32.3 \\
142  & 14.9  &  14.0  &  (23.0)\footnotemark[6], (23.6)\footnotemark[7] &  186 & 58.0 \\
\end{tabular}
\end{ruledtabular}
\footnotetext[1]{M. Schmidt and H. Haberland, Ref.\ \onlinecite{haberlandCR}.}
\footnotetext[2]{F. Calvo and F. Spiegelmann, Ref.\ \onlinecite{cs}.}
\footnotetext[3]{K. Manninen \textit{et al.}, Ref.\ \onlinecite{manninen04}.}
\footnotetext[4]{A. Aguado \textit{et al.},   Ref.\ \onlinecite{aguado-nalarge}.}
\footnotetext[5]{$N = 93$}
\footnotetext[6]{$N = 139$}
\footnotetext[7]{$N = 147$}
\end{table}

The KS melting temperatures $T_{m}$ are given along with the
approximate ``latent heats'' $L$ in Table \ref{table.data}.  
Following the convention in the experimental
works,\cite{haberland0,haberlandCR} we define $T_{m}$ here as the
maximum of the peak in the specific--heat.  We have also calculated a
quantity $\delta E$ defined as the average potential--energy of the
just melted cluster with respect to the ground--state structure at
$T=0$~K. Schmidt \emph{et al.}\cite{schmidt-na-dE} have inferred from
the experimental caloric curve that the melting temperature is
strongly influenced by such an energy contribution.  They showed
further that $T_m$ follows closely the variation in $\delta E$ as a
function of the cluster size.  In Table \ref{table.data}, we have also
given $T_m$ and $L$ calculated by the SMA parametrized
potential.\cite{manninen04,cs} The striking feature of the results
summarized in this table is that only the first principles KS--based
calculations reproduce, qualitatively as well as quantitatively, the
experimentally observed variation in the melting temperatures and the
latent heat for the three clusters investigated.\cite{comment1}

A noteworthy feature of the data in Table \ref{table.data} is that
Na$_{92}$ melts at a significantly lower temperature than Na$_{142}$.
This is true both for the KS data, which include electronic shell
effects explicitly, and for the SMA, DB, and ETF\cite{bb01} data,
which do not. We are thus led to conclude that geometry plays a
significant role in the melting--point depression of Na$_{92}$:
Na$_{55}$ and Na$_{142}$ are complete, or close to complete, Mackay
icosahedra, while Na$_{92}$ has surface growth on a two--shell
icosahedral core and is less stable.  However, as we shall see, the
electronic structure does play a subtle role in the behavior of the
melting temperatures of these clusters.

As mentioned earlier, previous simulations gave a melting temperature
for Na$_{55}$ significantly lower than the experimental one. Moreover,
for each of the SMA,\cite{manninen04,cs} ETF,\cite{bb01} and
DB\cite{aguado-nalarge} models, $T_{m}(55)$ is found to be
significantly less than $T_{m}(142)$ calculated within the same model,
while in the experiment $T_{m}(55)$ is very slightly greater than
$T_{m}(142)$.\cite{walesberry90} This discrepancy is largely removed
within the present KS model: we find that $T_{m}(55)\approx
T_{m}(142)$, within statistical error, and that both melting
temperatures agree with experiment.  This suggests quite strongly that
the high melting point of Na$_{55}$ relative to Na$_{142}$ is due to
electronic shell effects, since these are the main new element in our
KS approach not included in previous simulations.  This is
particularly clear in the case of the DB and ETF methods.  In these
approaches, quantum shell effects are effectively averaged out as a
function of size by the use of a density--dependent electron
kinetic--energy functional, and the total cluster energy follows quite
closely a smooth dependence on cluster size given by a
liquid--drop--type formula.\cite{bb01} In all other respects, however,
such as the use of pseudopotentials and density--dependent
exchange--correlation functionals, the DB and ETF methods are similar
to the present KS method.\cite{dbpseudo} We note that both Na$_{55}$
and Na$_{142}$ are close to ``magic'' systems~(Na$_{58}$ and
Na$_{138}$, respectively), but shell effects are relatively more
important for smaller systems,\cite{bohrmottelson} which could lead to
a relative enhancement in stability for Na$_{55}$ and thus a
relatively higher melting temperature.

\begin{figure}
{\centerline {\epsfxsize=0.6\textwidth  \epsffile{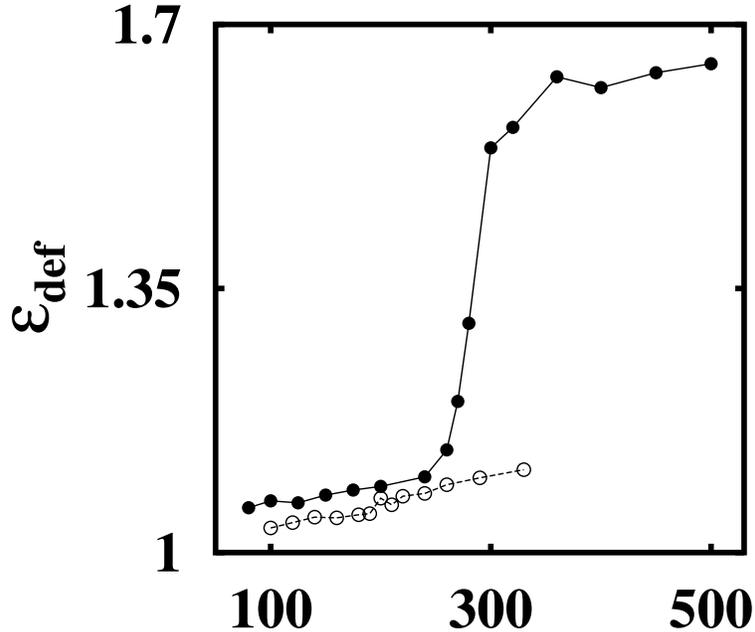}}}
{\caption{
\label{fig.eps-pro}
Time--averaged coefficient $\epsilon_{\rm def}$~(see text) describing
the degree of quadrupole deformation of Na$_{55}$~(continuous line)
and Na$_{92}$~(dotted line).
}}
\end{figure}

Further evidence for the role of quantum shell effects in melting may
be obtained by examining the \emph{shape} of the cluster before and
after melting.  In Fig.~\ref{fig.eps-pro} we plot the deformation
parameter $\epsilon_{\mathrm{def}}$ for Na$_{55}$, defined as
$\epsilon_{\mathrm{def}}={2Q_{1}}/(Q_{2}+Q_{3})$, where $Q_{1} \geq
Q_{2} \geq Q_{3}$ are the eigenvalues, in descending order, of the
quadrupole tensor $Q_{ij}=\sum_{I} R_{Ii}R_{Ij}.$ Here $i$ and $j$ run
from 1 to 3, $I$ runs over the number of ions, and $R_{Ii}$ is the
$i$th coordinate of ion $I$ relative to the cluster center of mass.  
A spherical system~($Q_{1} = Q_{2} = Q_{3}$) has
$\epsilon_{\mathrm{def}}=1$; a value $\epsilon_{\mathrm{def}}>1$
indicates a quadrupole deformation of some type.  From
Fig.~\ref{fig.eps-pro} we see that at low temperatures
$\epsilon_{\mathrm{def}}\approx 1$, corresponding to the compact
icosahedral ground--state, but as the cluster melts, the system
acquires a quadrupole deformation with $\epsilon_{\mathrm{def}}\approx
1.6$.  A more detailed investigation of this deformation with
two--dimensional deformation parameters such as the Hill--Wheeler
parameters\cite{bohrmottelson}~(not shown) indicates the cluster to be
undergoing shape fluctuations around a net prolate deformation.  A
typical deformed liquidlike structure is shown in Fig.\
\ref{fig.struc}(d).  A related phenomenon was observed earlier by
Rytk\"onen \emph{et al.}\cite{rytkonen98} for Na$_{40}$, except that
the magic Na$_{40}$ cluster underwent only an \emph{octupole}
deformation rather than a quadrupole deformation as observed here for
the nonmagic Na$_{55}$.  We do not observe statistically significant
quadrupole deformations for Na$_{92}$ or Na$_{142}$, for which
$\epsilon_{\mathrm{def}}\approx 1$ at all temperatures.

Interestingly, simulations of Na$_{55}$ carried out by us within the
SMA model do not show a deformation upon melting, but rather the
cluster remains essentially spherical at all temperatures,
$\epsilon_{\mathrm{def}}\approx 1$.  Since the SMA model explicitly
excludes quantum shell effects, we believe that the deformation of
Na$_{55}$ is due to the quantal Jahn--Teller distortion of the
open--shell system of valence electrons.  As further support for this
explanation, we note that the mean prolate deformation in the
liquidlike state agrees well with that found in jellium--model
calculations for neutral Na$_{55}$ that we have carried out, which
yield a uniaxial prolate deformation with a major-- to minor--axis
ratio of $R_{\mathrm{maj}}/R_{\mathrm{min}} \approx
\sqrt{\epsilon_{\mathrm{def}}} \approx 1.3$, in close agreement with
the liquidlike simulations. Evidently, the compact ground--state
structure is favored by the possibility of geometric packing into an
icosahedron, while in the nonrigid liquidlike state the cluster can
lower its free energy by undergoing a spontaneous shape deformation.  
On this reasoning, the magic Na$_{92}$ cluster would not be expected
to deform upon melting, consistent with the observation.  On the other
hand, the Na$_{142}$ cluster \emph{would} be expected to deform, at
least slightly; presumably, the Jahn--Teller forces here are
sufficiently weak that any deformation is not statistically
significant.

Note that, as mentioned earlier, the melting temperature is strongly
influenced by the potential--energy difference $\delta E$ between
liquidlike and solidlike states.  Therefore, even if the ground--state
is quite spherical, as for Na$_{55}$, the melting temperature may
still be influenced by important quantal deformation effects entering
only in the liquidlike state.  The KS simulations undertaken here
incorporate all these various effects correctly.

\section{Conclusion}

In conclusion, the KS approach appears to be capable of making
quantitative predictions of melting temperatures and latent heats in
Na clusters.  Na$_{55}$, Na$_{92}$, and Na$_{142}$ are each magic or
nearly magic, but only Na$_{55}$ and Na$_{142}$ are also close to
icosahedral shell closures.  The fact that Na$_{92}$ melts at a
significantly lower temperature than the other two shows that
geometric effects are very important in determining the pattern of
melting temperatures observed experimentally.  However, electronic
shell effects can play an important role too, both in influencing
overall binding energies and bond--lengths as a function of size, and
indirectly via shape deformation effects that may arise differently in
the solidlike and liquidlike states.  For an accurate treatment of the
metallic bonding, and a quantitative prediction of melting
temperatures and latent heats, it is essential to incorporate
electronic shell effects appropriately, as is possible, for example,
within the KS approach used here.  This is especially true for the
smaller sizes of clusters. There is a size regime up to $N \approx
150$ or so where a full KS treatment is warranted.

\section{Acknowledgment}

One of us (SC) acknowledges financial support from the Center for
Modeling and Simulation, University of Pune. We gratefully acknowledge
the support of the Indo--French Center for Promotion for Advance
Research. It is a pleasure to acknowledge C--DAC~(Pune) for providing
us with supercomputing facilities.


\begin{references}
\bibitem{aguado-nalarge}
  A. Aguado, J. M. L\'{o}pez, J. A. Alonso, and M. J. Stott, 
  J. Phys. Chem. B \textbf{105}, 2386 (2001). 
  The $T_m$ for Na$_{92}$ noted in this reference is for the
  icosahedral isomer, which is not the correct ground--state.
\bibitem{deltarms}
   $\delta_{\mathrm{rms}}$ is defined in Ref.~\onlinecite{snmelting},
   for example, and corresponds to a finite--sized version of the bulk
   Lindemann criterion.
\bibitem{vasp-wfn-extrapol}
In the \textsc{vasp} package during molecular--dynamics, a
second--order extrapolation of the wavefunctions and charge density is
performed.  This reduces the number of self--consistency iterations
required to as few as 4--6.
\bibitem{comment1}
   Although our calculations are for neutral clusters,
   we do not expect the results to change significantly for
   the singly charged clusters studied experimentally. 
   See Ref.~\onlinecite{cs} for a model.
\bibitem{dbpseudo}
   There are also differences between the pseudopotentials used in the
   DB and ETF calculations and in the present KS calculation.  However,
   the DB calculation, Ref.~\onlinecite{aguado-nalarge}, used a
   norm--conserving pseudopotential, while the ETF calculation,
   Ref.~\onlinecite{bb01}, used a softer, phenomenological
   pseudopotential, and in both cases it was found that $T_m(55)$ was
   significantly less than $T_m(142)$, suggesting that this result is
   largely independent of the pseudopotential used.  In the present KS
   calculations, as we have mentioned, the ultrasoft pseudopotentials
   compare well with (nearly) all--electron PAW calculations.  We
   therefore believe that differences in the pseudopotential are unlikely
   to account for the success of the present KS method in reproducing the
   high melting point of $T_m(55)$ relative to $T_m(142)$.
\end{references}
\end{document}